\pdfoutput=1
\documentclass[12pt,a4paper,]{article}

\usepackage{ifthen} 
\newboolean{pdflatex}
\setboolean{pdflatex}{true} 

\newboolean{articletitles}
\setboolean{articletitles}{true} 

\newboolean{uprightparticles}
\setboolean{uprightparticles}{false} 

\def\paperauthors{LHCb collaboration} 
\def\paperasciititle{Search for the doubly heavy baryon Xibc+ ->
  J/psi Xic+} 
\def\papertitle{Search for the doubly heavy baryon $\Xibcp$
  decaying to
  $\jpsi \Xicp$} 
\def\paperkeywords{{High Energy Physics}, {LHCb}} 
\def\papercopyright{\the\year\ CERN for the benefit of the LHCb collaboration} 
\def\paperlicence{CC BY 4.0 licence}
\def\paperlicenceurl{https://creativecommons.org/licenses/by/4.0/}

\usepackage[top=1in, bottom=1.25in, left=1in, right=1in]{geometry}

\usepackage{microtype}
\usepackage{lineno}  
\usepackage{xspace} 
\usepackage{caption} 

\usepackage{graphicx}  
\usepackage{color}
\usepackage{colortbl}
\graphicspath{{./figs/}} 

\usepackage{amsmath} 
\usepackage{amssymb}
\usepackage{amsfonts}
\usepackage{upgreek} 

\newcommand*\patchAmsMathEnvironmentForLineno[1]{%
\expandafter\let\csname old#1\expandafter\endcsname\csname #1\endcsname
\expandafter\let\csname oldend#1\expandafter\endcsname\csname
end#1\endcsname
 \renewenvironment{#1}%
   {\linenomath\csname old#1\endcsname}%
   {\csname oldend#1\endcsname\endlinenomath}%
}
\newcommand*\patchBothAmsMathEnvironmentsForLineno[1]{%
  \patchAmsMathEnvironmentForLineno{#1}%
  \patchAmsMathEnvironmentForLineno{#1*}%
}
\AtBeginDocument{%
\patchBothAmsMathEnvironmentsForLineno{equation}%
\patchBothAmsMathEnvironmentsForLineno{align}%
\patchBothAmsMathEnvironmentsForLineno{flalign}%
\patchBothAmsMathEnvironmentsForLineno{alignat}%
\patchBothAmsMathEnvironmentsForLineno{gather}%
\patchBothAmsMathEnvironmentsForLineno{multline}%
\patchBothAmsMathEnvironmentsForLineno{eqnarray}%
}

\usepackage{hyperxmp}

\usepackage[pdftex,
            pdfauthor={\paperauthors},
            pdftitle={\paperasciititle},
            pdfkeywords={\paperkeywords},
            pdfcopyright={Copyright (C) \papercopyright},
            pdflicenseurl={\paperlicenceurl}]{hyperref}

\usepackage[colorinlistoftodos,textsize=scriptsize]{todonotes}

\usepackage[bottom,flushmargin,hang,multiple]{footmisc}

\usepackage[all]{hypcap} 

\usepackage{xspace} 
\usepackage{upgreek}

\def\lhcb   {\mbox{LHCb}\xspace}

\def\MagUp {\mbox{\em Mag\kern -0.05em Up}\xspace}

\ifthenelse{\boolean{uprightparticles}}%
{

 \def\Pmu         {\ensuremath{\upmu}\xspace}

 \def\Ppsi        {\ensuremath{\uppsi}\xspace}

 \def\PDelta      {\ensuremath{\Delta}\xspace}                 
 \def\PXi         {\ensuremath{\Xi}\xspace}                 
 \def\PLambda     {\ensuremath{\Lambda}\xspace}                 
 \def\PSigma      {\ensuremath{\Sigma}\xspace}                 
 \def\POmega      {\ensuremath{\Omega}\xspace}                 
 \def\PUpsilon    {\ensuremath{\Upsilon}\xspace}
 \let\oldPi\Pi
 \def\PPi         {\ensuremath{\oldPi}\xspace}

 \def\PB      {\ensuremath{\mathrm{B}}\xspace}                 
                  
 \def\PD      {\ensuremath{\mathrm{D}}\xspace}

 \def\PJ      {\ensuremath{\mathrm{J}}\xspace}                 
 \def\PK      {\ensuremath{\mathrm{K}}\xspace}

 \def\Pb      {\ensuremath{\mathrm{b}}\xspace}                 
 \def\Pc      {\ensuremath{\mathrm{c}}\xspace}

 \def\Pi      {\ensuremath{\mathrm{i}}\xspace}

 \def\Ps      {\ensuremath{\mathrm{s}}\xspace}

 \def\thebaroffset{0.0em}
}
{

 \def\Pmu         {\ensuremath{\mu}\xspace}

 \def\Ppsi        {\ensuremath{\psi}\xspace}                 
                  
 \mathchardef\PDelta="7101
 \mathchardef\PXi="7104
 \mathchardef\PLambda="7103
 \mathchardef\PSigma="7106
 \mathchardef\POmega="710A
 \mathchardef\PUpsilon="7107
 \mathchardef\PPi="7105
                  
 \def\PB      {\ensuremath{B}\xspace}                 
                  
 \def\PD      {\ensuremath{D}\xspace}

 \def\PJ      {\ensuremath{J}\xspace}                 
 \def\PK      {\ensuremath{K}\xspace}

 \def\Pb      {\ensuremath{b}\xspace}                 
 \def\Pc      {\ensuremath{c}\xspace}

 \def\Pi      {\ensuremath{i}\xspace}

 \def\Ps      {\ensuremath{s}\xspace}

 \def\thebaroffset{0.18em}
}
\newcommand{\offsetoverline}[2][\thebaroffset]{\kern #1\overline{\kern -#1 #2}}%

\makeatletter
\ifcase \@ptsize \relax
  \newcommand{\miniscule}{\@setfontsize\miniscule{4}{5}}
\or
  \newcommand{\miniscule}{\@setfontsize\miniscule{5}{6}}
\or
  \newcommand{\miniscule}{\@setfontsize\miniscule{5}{6}}
\fi
\makeatother

\DeclareRobustCommand{\optbar}[1]{\shortstack{{\miniscule (\rule[.5ex]{1.25em}{.18mm})}
  \\ [-.7ex] $#1$}}

\def\mup        {{\ensuremath{\Pmu^+}}\xspace}
\def\mun        {{\ensuremath{\Pmu^-}}\xspace} 

\def\squark    {{\ensuremath{\Ps}}\xspace}

\def\cquark    {{\ensuremath{\Pc}}\xspace}

\def\bquark    {{\ensuremath{\Pb}}\xspace}

\def\KorKbar {\kern \thebaroffset\optbar{\kern -\thebaroffset \PK}{}\xspace}

\def\D       {{\ensuremath{\PD}}\xspace}

\def\DorDbar {\kern \thebaroffset\optbar{\kern -\thebaroffset \PD}\xspace}

\def\Dp      {{\ensuremath{\D^+}}\xspace}
\def\Dm      {{\ensuremath{\D^-}}\xspace}

\def\DpDm    {\ensuremath{\Dp {\kern -0.16em \Dm}}\xspace}

\def\Ds      {{\ensuremath{\D^+_\squark}}\xspace}
\def\Dsp     {{\ensuremath{\D^+_\squark}}\xspace}

\def\B       {{\ensuremath{\PB}}\xspace}

\def\BorBbar {\kern \thebaroffset\optbar{\kern -\thebaroffset \PB}\xspace}

\def\Bd      {{\ensuremath{\B^0}}\xspace}

\def\BdorBdbar {\kern \thebaroffset\optbar{\kern -\thebaroffset \Bd}\xspace}

\def\Bs      {{\ensuremath{\B^0_\squark}}\xspace}

\def\BsorBsbar {\kern \thebaroffset\optbar{\kern -\thebaroffset \Bs}\xspace}

\def\Bcp     {{\ensuremath{\B_\cquark^+}}\xspace}

\def\jpsi     {{\ensuremath{{\PJ\mskip -3mu/\mskip -2mu\Ppsi}}}\xspace}

\def\Y#1S{\ensuremath{\PUpsilon{(#1S)}}\xspace}

\def\LorLbar     {\kern \thebaroffset\optbar{\kern -\thebaroffset \PLambda}\xspace}

\def\Xires       {{\ensuremath{\PXi}}\xspace}

\def\Omegares    {{\ensuremath{\POmega}}\xspace}

\def\Xicp        {{\ensuremath{\Xires^+_\cquark}}\xspace}

\def\BF         {{\ensuremath{\mathcal{B}}}\xspace}
\def\BR         {\BF}

\def\to                 {\ensuremath{\rightarrow}\xspace}

\def\eps   {{\ensuremath{\varepsilon}}\xspace}

\def\AT#1     {\ensuremath{A_{\mathrm{T}}^{#1}}\xspace}           

\def\C#1      {\ensuremath{\mathcal{C}_{#1}}\xspace}                       
\def\Cp#1     {\ensuremath{\mathcal{C}_{#1}^{'}}\xspace}                    
\def\Ceff#1   {\ensuremath{\mathcal{C}_{#1}^{\mathrm{(eff)}}}\xspace}        
\def\Cpeff#1  {\ensuremath{\mathcal{C}_{#1}^{'\mathrm{(eff)}}}\xspace}       
\def\Ope#1    {\ensuremath{\mathcal{O}_{#1}}\xspace}                       
\def\Opep#1   {\ensuremath{\mathcal{O}_{#1}^{'}}\xspace}                    

\newcommand{\nospaceunit}[1]{\ensuremath{\text{#1}}}       
\newcommand{\aunit}[1]{\ensuremath{\text{\,#1}}}       

\newcommand{\tev}{\aunit{Te\kern -0.1em V}\xspace}
\newcommand{\gev}{\aunit{Ge\kern -0.1em V}\xspace}
\newcommand{\mev}{\aunit{Me\kern -0.1em V}\xspace}
\newcommand{\kev}{\aunit{ke\kern -0.1em V}\xspace}
\newcommand{\ev}{\aunit{e\kern -0.1em V}\xspace}
 
\newcommand{\mevc}{\ensuremath{\aunit{Me\kern -0.1em V\!/}c}\xspace}
\newcommand{\gevc}{\ensuremath{\aunit{Ge\kern -0.1em V\!/}c}\xspace}
\newcommand{\mevcc}{\ensuremath{\aunit{Me\kern -0.1em V\!/}c^2}\xspace}
\newcommand{\gevcc}{\ensuremath{\aunit{Ge\kern -0.1em V\!/}c^2}\xspace}

\def\mum  {\ensuremath{\,\upmu\nospaceunit{m}}\xspace}

\def\fb   {\ensuremath{\aunit{fb}}\xspace}
\def\invfb   {\ensuremath{\fb^{-1}}\xspace}

\def\fs   {\aunit{fs}}

\newcommand{\chisq}{\ensuremath{\chi^2}\xspace}

\newcommand{\chisqip}{\ensuremath{\chi^2_{\text{IP}}}\xspace}

\def\gsim{{~\raise.15em\hbox{$>$}\kern-.85em
          \lower.35em\hbox{$\sim$}~}\xspace}
\def\lsim{{~\raise.15em\hbox{$<$}\kern-.85em
          \lower.35em\hbox{$\sim$}~}\xspace}

\def\sqs   {\ensuremath{\protect\sqrt{s}}\xspace}

\def\pt         {\ensuremath{p_{\mathrm{T}}}\xspace}

\def\ptot       {\ensuremath{p}\xspace}

\def\evtgen     {\mbox{\textsc{EvtGen}}\xspace}

\def\geant      {\mbox{\textsc{Geant4}}\xspace}

\def\photos     {\mbox{\textsc{Photos}}\xspace}

\def\pythia     {\mbox{\textsc{Pythia}}\xspace}

\def\tell1  {TELL1\xspace}
\def\ukl1   {UKL1\xspace}

\newcommand{\lhcborcid}[1]{\href{https://orcid.org/#1}{\hspace*{0.1em}\raisebox{-0.45ex}{\includegraphics[width=1em]{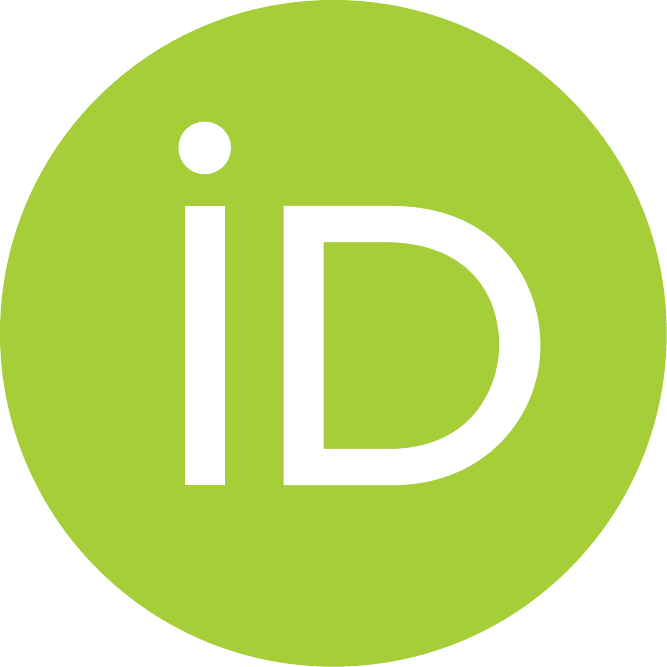}}}}

\def\Xibcp       {{\ensuremath{\Xires^+_{\bquark\cquark}}}\xspace}
\def\Xibcz       {{\ensuremath{\Xires^0_{\bquark\cquark}}}\xspace}
\def\Omegabcz       {{\ensuremath{\Omegares^0_{\bquark\cquark}}}\xspace}

\usepackage{cite} 
\usepackage{mciteplus}

\usepackage{longtable} 

\begin{document}

\renewcommand{\thefootnote}{\fnsymbol{footnote}}
\setcounter{footnote}{1}


\begin{titlepage}
\pagenumbering{roman}

\vspace*{-1.5cm}
\centerline{\large EUROPEAN ORGANIZATION FOR NUCLEAR RESEARCH (CERN)}
\vspace*{1.5cm}
\noindent
\begin{tabular*}{\linewidth}{lc@{\extracolsep{\fill}}r@{\extracolsep{0pt}}}
\ifthenelse{\boolean{pdflatex}}
{\vspace*{-1.5cm}\mbox{\!\!\!\includegraphics[width=.14\textwidth]{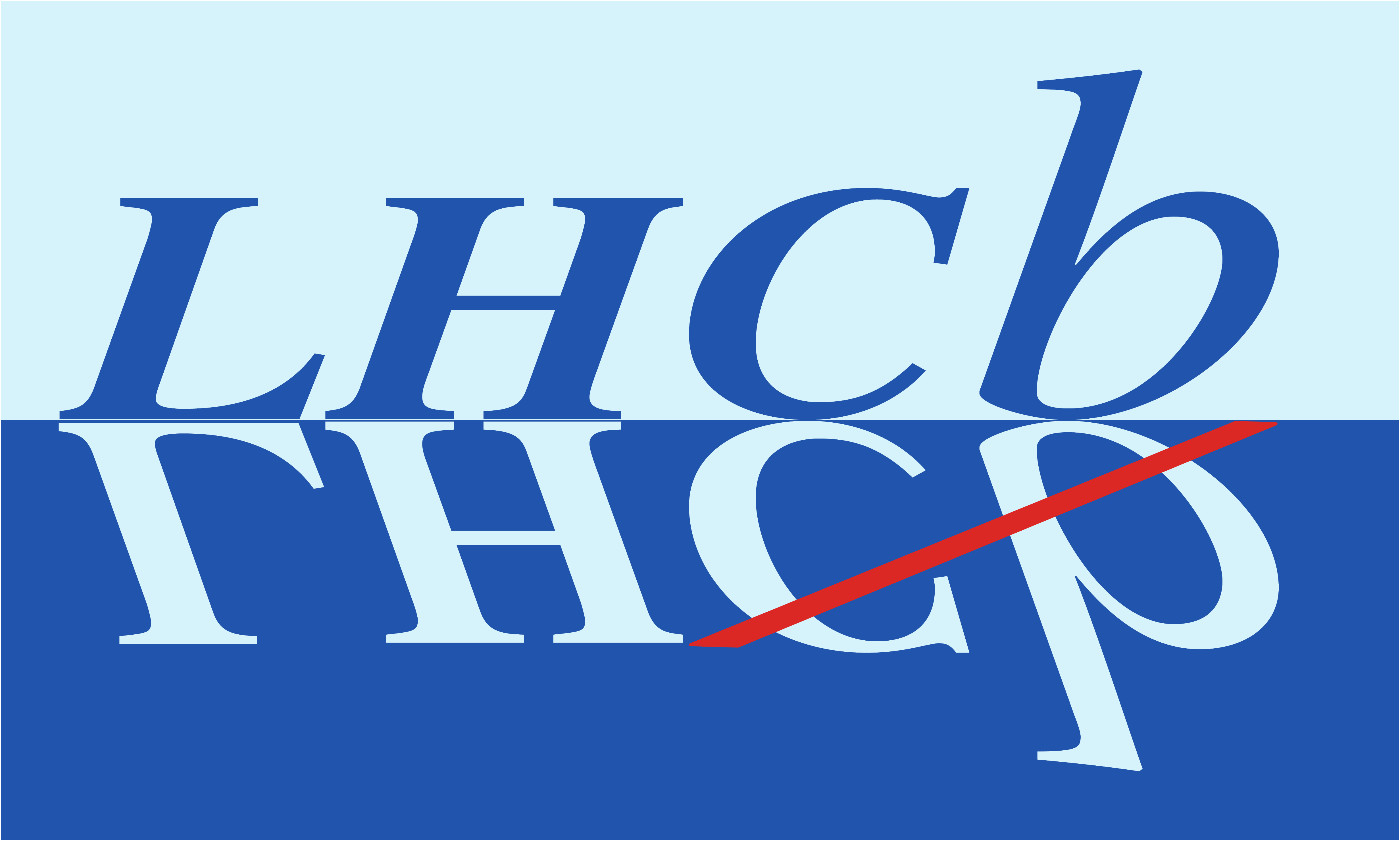}} & &}%
{\vspace*{-1.2cm}\mbox{\!\!\!\includegraphics[width=.12\textwidth]{figs/lhcb-logo.eps}} & &}%
\\
 & & CERN-EP-2022-048 \\  
 & & LHCb-PAPER-2022-005 \\  
 & & 24 July 2023 \\ 
 & &  \\
\end{tabular*}

\vspace*{4.0cm}

{\normalfont\bfseries\boldmath\huge
\begin{center}
  \papertitle 
\end{center}
}

\vspace*{2.0cm}

\begin{center}
\paperauthors\footnote{Authors are listed at the end of this paper.} 
\end{center}

\vspace{\fill}

\begin{abstract}
  \noindent
A first search for the $\Xibcp\to\jpsi\Xicp$ decay is performed by the LHCb experiment with a data sample of proton-proton collisions, corresponding to an integrated luminosity of 9\invfb recorded at centre-of-mass energies of 7, 8, and 13\tev.
Two peaking structures are seen with a local (global) significance of $4.3\,(2.8)$ and $4.1\,(2.4)$ standard deviations at masses of $6571\mevcc$ and $6694\mevcc$, respectively. 
Upper limits are set on the \Xibcp baryon production cross-section 
times the branching fraction relative to that of the $\Bcp\to\jpsi\Dsp$ decay at centre-of-mass energies of 8 and 13\tev, in the $\Xibcp$ and in the $\Bcp$ rapidity and transverse-momentum ranges from 2.0 to 4.5 and 0 to $20\gevc$, respectively. Upper limits are presented as a function of the \mbox{$\Xibcp$ mass} and lifetime. 
\end{abstract}

\vspace*{2.0cm}

\begin{center}
  Published in Chin. Phys. C 47 (2023) 093001
\end{center}

\vspace{\fill}

{\footnotesize 
\centerline{\copyright~\papercopyright. \href{\paperlicenceurl}{\paperlicence}.}}
\vspace*{2mm}

\end{titlepage}


\newpage
\setcounter{page}{2}
\mbox{~}


\renewcommand{\thefootnote}{\arabic{footnote}}
\setcounter{footnote}{0}

\cleardoublepage


\pagestyle{plain} 
\setcounter{page}{1}
\pagenumbering{arabic}


\section{Introduction}
\label{sec:intro}

Doubly heavy baryons consisting of two heavy quarks ($b$ or $c$) and one light quark ($u$, $d$, or $s$) are expected within the 
quark model~\cite{GellMann:1964nj,Zweig:352337, *Zweig:570209}.
In proton-proton ($pp$) collisions at the Large Hadron Collider, 
a possible model for production of these states is through gluon-gluon fusion, 
$g + g \to Q_1\overline{Q}_1 + Q_2\overline{Q}_2$ ($Q$ denotes a heavy quark), a process that can be computed using perturbative quantum chromodynamics (QCD)~\cite{Ma:2003zk,Chang:2006xp,Chang:2009va}. 
The doubly heavy baryon is then formed via hadronisation where the two heavy quarks form a diquark which binds with a light quark. Other models exist, including production at the scale of the hard process, or production via non-perturbative effects such as colour reconnection.
The measurement of the properties of these doubly heavy baryons provides insight into both their production mechanism and internal structure.

The observation and properties of the doubly heavy $\PXi_{cc}^{++}$ ($ccu$) baryon have been firmly established by the LHCb collaboration~\cite{LHCb-PAPER-2017-018,LHCb-PAPER-2018-019,LHCb-PAPER-2019-011,LHCb-PAPER-2019-035,LHCb-PAPER-2019-037}, while the $\PXi_{cc}^{+}$ ($ccd$) and $\POmega_{cc}^+$ ($ccs$) baryons have been searched for~\cite{LHCb-PAPER-2019-029,LHCb-PAPER-2021-019,LHCb-PAPER-2021-011}, and only hints of a signal were seen. The LHCb collaboration has also carried out searches for the neutral doubly heavy baryons, $\Xibcz~(bcd)$~\cite{LHCb-PAPER-2020-014} and $\Omegabcz~(bcs)$~\cite{LHCb-PAPER-2021-002}, but these states are yet to be observed.

To date, no search has been performed for the $\Xibcp$ baryon, a bound state with quark content $bcu$. This baryon is expected to have a mass in the range of \mbox{6700--7029\mevcc~\cite{Ponce:1978gk,Roncaglia:1995az,Silvestre-Brac:1996tmn,Lichtenberg:1995kg,Ebert:1996ec,Kiselev:2001fw,Ebert:2002ig,He:2004px,Albertus:2006ya,Roberts:2007ni,Zhang:2008rt,Gerasyuta:2008zy,Giannuzzi:2009gh,Weng:2010rb,Aliev:2012ru,Karliner:2014gca,Brown:2014ena,Weng:2018mmf,Li:2021bkj,Wang:2022ufh}}, while its lifetime is
predicted to be 
between $240\fs$ and $607\fs$~\cite{Kiselev:1999kh,Kiselev:2001fw,Karliner:2014gca,Berezhnoy:2018bde,Cheng:2019sxr}. The $\Xibcp$ production cross-section at a centre-of-mass energy of $\sqs =13\tev$ is predicted to be about 16~nb~\cite{Zhang:2011hi}  in the fiducial region $\pt>4\gevc$ and $1.9<\eta<4.9$, where $\pt$ is the momentum component transverse to the beam direction and $\eta$ is the pseudorapidity. 

This article presents the first search for the $\Xibcp$ baryon through its decay via the $\jpsi \Xicp$ channel, with $\jpsi \to \mu^{+}\mu^{-}$ and $\Xicp \to p K^{-}\pi^{+}$ final states, using $pp$ collision data collected by the LHCb experiment at centre-of-mass energies of 7, 8, and 13\tev, corresponding to integrated luminosities of 1, 2, and 6\invfb, respectively. 

\begin{figure}[!b]
  \begin{center}
    \resizebox{0.5\textwidth}{!}{%
      \includegraphics{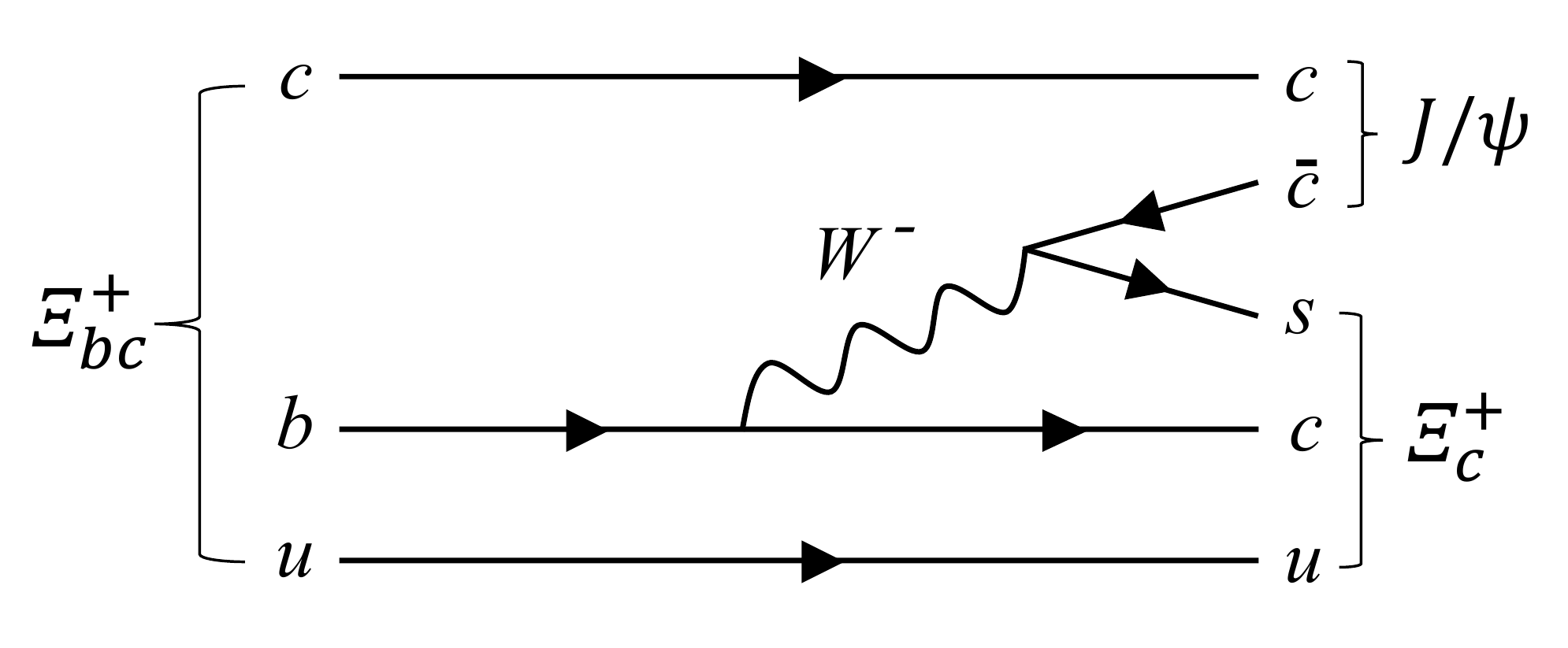}}
   \caption{Example leading-order Feynman diagram for the $\Xibcp \to \jpsi \Xicp$ decay.}
   \label{fig:diagram}
 \end{center} 
\end{figure}

The search for this decay is advantageous over the previous $\Xibcz$ searches in several ways. First, the $\Xibcp$ baryon is expected to have a larger lifetime than that of the $\Xibcz$ baryon~\cite{Kiselev:1999kh,Kiselev:2001fw,Karliner:2014gca,Berezhnoy:2018bde,Cheng:2019sxr}, which leads to a larger selection efficiency as the lifetime information is used to suppress background from primary $pp$ interactions. Second, the mode studied here uses $\jpsi\to\mup\mun$ decays, which typically have a selection efficiency three times larger  than the fully hadronic modes used in the previous $\Xibcz$ searches. Last, the modes used in the $\Xibcz$ analyses involved suppressed $b\to u$ or 
$b\to s$ transitions, or $W$-exchange diagrams. 
Here the decay to  the $\jpsi\Xicp$ final state involves a colour-suppressed $b\to c$ transition, with a decay amplitude that is less likely to be suppressed, as shown in Fig.~\ref{fig:diagram}.

To reduce systematic uncertainties, the \Xibcp production cross-section times the $\mbox{\Xibcp \to \jpsi \Xicp}$ branching fraction is measured relative to that of the normalisation mode $\mbox{\Bcp \to \jpsi \Dsp}$ with $\jpsi \to \mu^{+}\mu^{-}$ and $\Dsp \to K^{+} K^{-}\pi^{+}$ decays. Specifically, the quantity $\mathcal{R}$ is defined as
\begin{equation}
\label{eq:prodratio}
\mathcal{R}=\frac{\sigma(\Xibcp)\times\BF(\Xibcp \to \jpsi \Xicp)\times \BF(\Xicp \to pK^-\pi^+)}{\sigma(B_c^+)\times\BF(B_c^+ \to \jpsi D_s^+)\times \BF(D_s^+ \to K^+K^-\pi^+)},\end{equation}
where $\sigma(\Xibcp)$ and $\sigma(B_c^+)$ are the production cross-sections of $\Xibcp$ and $\Bcp$ hadrons, respectively, 
and $\BR$ is the branching fraction of the corresponding decay.
The ratio $\mathcal{R}$ is measured in the rapidity range $2.0 < y < 4.5$ and in the \pt region from 0 to 20$\gevc$. Measurements of ${\mathcal{R}}$ are reported for the $\sqs=8$ and 13\tev data sets collected in 2012 and 2016--2018, corresponding to integrated luminosities of 2 and 5.4\invfb, respectively. The 2011 data sample taken with a centre-of-mass 
energy of 7\tev is small, and it is not used in the production rate measurement. 

The ratio $\mathcal{R}$ is evaluated as
\begin{equation}
\label{eq:prodratio1}
\mathcal{R}=\frac{\varepsilon_{\rm norm}}{\varepsilon_{\rm sig}}\frac{N_{\rm sig}}{N_{\rm norm}}\equiv \alpha N_{\rm sig},
\end{equation}
where $\varepsilon_{\rm sig}$ and $\varepsilon_{\rm norm}$ are the total efficiencies of the $\Xibcp$ signal and $\Bcp$ normalisation decay modes,
$N_{\rm sig}$ and $N_{\rm norm}$ are the corresponding signal yields, and the derived quantity $\alpha$ is the single-event sensitivity. 

An estimate for $\mathcal{R}$ can be obtained by assuming that the ratio of production cross-sections $\sigma(\Xibcp)/\sigma(\Bcp)$ is about 0.4~\cite{Chang:2003cr,Gao:2010zzc,Zhang:2011hi}, $\BF(\Xibcp \to \jpsi \Xicp)\sim 1/3\cdot\BF(\Bcp \to \jpsi \Dsp)$ due to colour suppression, $\BF(\Xicp \to pK^-\pi^+)=(0.62\pm0.30)\%$~\cite{Li:2019atu,LHCb-PAPER-2020-016}, $\BF(\Dsp \to K^+K^-\pi^+)=(5.39\pm0.15)\%$~\cite{PDG2020}, and assuming an efficiency ratio 
$\varepsilon_{\rm sig}/\varepsilon_{\rm norm}\sim 1$. 
With these inputs, the value  $\mathcal{R}\sim0.015$ is obtained.
With 1100 $\Bcp \to \jpsi \Dsp$ candidates observed in the full data set collected by the LHCb experiment~\cite{LHCb-PAPER-2020-003}, approximately 15 reconstructed $\Xibcp \to \jpsi \Xicp$ signal decays are expected in the LHCb detector acceptance.

\section{LHCb detector and simulation}
\label{sec:det}

The \lhcb detector~\cite{LHCb-DP-2008-001,LHCb-DP-2014-002} is a single-arm forward
spectrometer covering the range $2<\eta <5$,
designed for the study of particles containing \bquark or \cquark
quarks. It includes a high-precision tracking system
consisting of a silicon-strip vertex detector surrounding the $pp$
interaction region~\cite{LHCb-DP-2014-001}, a large-area silicon-strip detector located
upstream of a dipole magnet with a bending power of about
$4{\mathrm{\,Tm}}$, and three stations of silicon-strip detectors and straw
drift tubes~\cite{LHCb-DP-2013-003,LHCb-DP-2017-001}
placed downstream of the magnet.
The tracking system provides a measurement of the momentum, \ptot, of charged particles with
a relative uncertainty that varies from 0.5\% at low \ptot to 1.0\% at 200\gevc.
The minimum distance of a track to a primary $pp$ collision vertex (PV), the impact parameter (IP), 
is measured with a resolution of $(15+29/\pt)\mum$,
where \pt is expressed in\,\gevc.
Different types of charged hadrons are distinguished using information
from two ring-imaging Cherenkov detectors~\cite{LHCb-DP-2012-003}. 
Photons, electrons and hadrons are identified by a calorimeter system consisting of
scintillating-pad and preshower detectors, an electromagnetic
and a hadronic calorimeter. Muons are identified by a
system composed of alternating layers of iron and multiwire
proportional chambers~\cite{LHCb-DP-2012-002}.
The online event selection is performed by a trigger~\cite{LHCb-DP-2012-004}, 
which consists of a hardware stage, based on information from the calorimeter and muon
systems,  followed by a two-level software stage, which applies a full event
reconstruction.

Simulation of signal and normalisation modes is required to model the effects of the detector acceptance and the imposed selection requirements.
  In the simulation, $pp$ collisions are generated using
  \pythia8~\cite{Sjostrand:2007gs,*Sjostrand:2006za} 
  with a specific \lhcb configuration~\cite{LHCb-PROC-2010-056}.
A dedicated generator, $\textsc{GenXicc}$~2.0~\cite{Chang:2009va}, is used to simulate the doubly heavy baryon production,
and is interfaced to \pythia8 to provide underlying event, parton shower, and hadronisation generation. 
Decays of unstable particles are described by \evtgen~\cite{Lange:2001uf}, in which final-state radiation is generated using \photos~\cite{davidson2015photos}. The interaction of the generated particles with the material of the detector, and its response, are simulated using the \geant toolkit \cite{Agostinelli:2002hh} as described in Ref.~\cite{LHCb-PROC-2011-006}. The simulated $\Xibcp$ events are generated with a mass of $6900\mevcc$ and a lifetime of $400\fs$, and samples with different mass and lifetime hypotheses are obtained using a weighting technique, described in Sec.~\ref{sec:res}. The $\Xibcp$ baryon is assumed to decay uniformly within the allowed phase space.

\section{Reconstruction and selection}
\label{sec:sel}
For both the $\Xibcp$ signal and the $\Bcp$ normalisation mode, $\jpsi$ candidates are reconstructed from $\mu^+ \mu^-$ pairs. In the online event selection the $\jpsi$ candidates are required to pass dedicated trigger requirements. At the hardware stage at least one high-$\pt$ muon is required. In the subsequent software stage, a pair of oppositely charged muon candidates is required to originate from a common vertex that is displaced from any PV and to have an invariant mass $m(\mu^+\mu^-)$ in a wide window around the $\jpsi$ mass.

In the offline selection, $\jpsi$ candidates are formed by two oppositely charged tracks identified as muons, with $m(\mu^+\mu^-)$ in the range between 3040 and 3140$\mevcc$, and that are required to be consistent with a common vertex that is significantly displaced from any PV. For the $\Xibcp$ signal mode, the $\Xicp$ candidates are reconstructed in the $p K^- \pi^+$ final state. The three tracks, identified as a proton, kaon, and pion, are required to be inconsistent with originating from any PV in the event and form a common vertex with good fit quality. The reconstructed mass of the $\Xicp$ candidates is required to be within $\pm 15\mevcc$ of the known $\Xicp$ mass~\cite{PDG2020}, which corresponds to about 2.5 times the mass resolution. For the $\Bcp$ normalisation mode, the $\Dsp$ candidates are reconstructed in the $K^+ K^- \pi^+$ final state, and selected in the same way as the $\Xicp$ candidates, apart from different particle identification (PID) requirements. The $\jpsi$ and $\Xicp\; (\Dsp)$ candidates are then required to form a common vertex with a good fit quality to reconstruct the $\Xibcp\; (\Bcp)$ candidate. Each $\Xibcp$ ($\Bcp$) is associated to the PV in the event for which $\chisqip$ is the smallest, where $\chisqip$ is the difference in $\chisq$ of the PV fit with and without the $\Xibcp$ ($\Bcp$) included in the PV fit. The $\Xibcp$ ($\Bcp$) candidate is required to have a trajectory that points back to its associated PV. 

A boosted decision tree (BDT) algorithm~\cite{Breiman,AdaBoost} implemented in the TMVA package~\cite{Hocker:2007ht,*TMVA4} is applied to both the signal and the normalisation candidates to further improve the purity of the samples. To train this classifier,
simulated $\Xibcp$ baryon decays are used as signal proxy and candidates lying in the upper $\jpsi\Xicp$ mass sideband, with $m(\jpsi\Xicp)$ in the range of 8000--8500\mevcc, are used to model the background. The BDT algorithm uses kinematic and vertex-topology variables that show a good discrimination power between signal and background samples. The variables include: the $\chisqip$, \ptot and \pt of all particles; PID variables for the final state particles; the flight-distance $\chisq$ between the PV and the decay vertex; and the vertex fit quality of the $\jpsi$, $\Xicp$ and $\Xibcp$ candidates. The flight-distance $\chisq$ is defined as the $\chisq$ of the hypothesis that the decay vertex of the candidate coincides with its associated PV. 

The threshold of the BDT response is set by maximising the figure of merit~\cite{Punzi:2003bu} $\varepsilon/(\beta/2+\sqrt{N_B})$, where $\varepsilon$ is the signal efficiency estimated from simulation, $\beta$ corresponds to the number of standard deviations in a Gaussian significance test, which is taken as 5, and $N_B$ is the number of background candidates determined in the upper sideband and extrapolated to the signal region. The performance of the BDT classifier is tested and found to be stable against the $\Xibcp$ lifetime in the range from 300 to $500\fs$.
The same BDT selection is applied to the normalisation mode, with the PID requirements changed accordingly.

\section{Yield measurements}
\label{Sec:Yield}
\begin{figure}
  \begin{center}
    \resizebox{0.6\textwidth}{!}{%
      \includegraphics{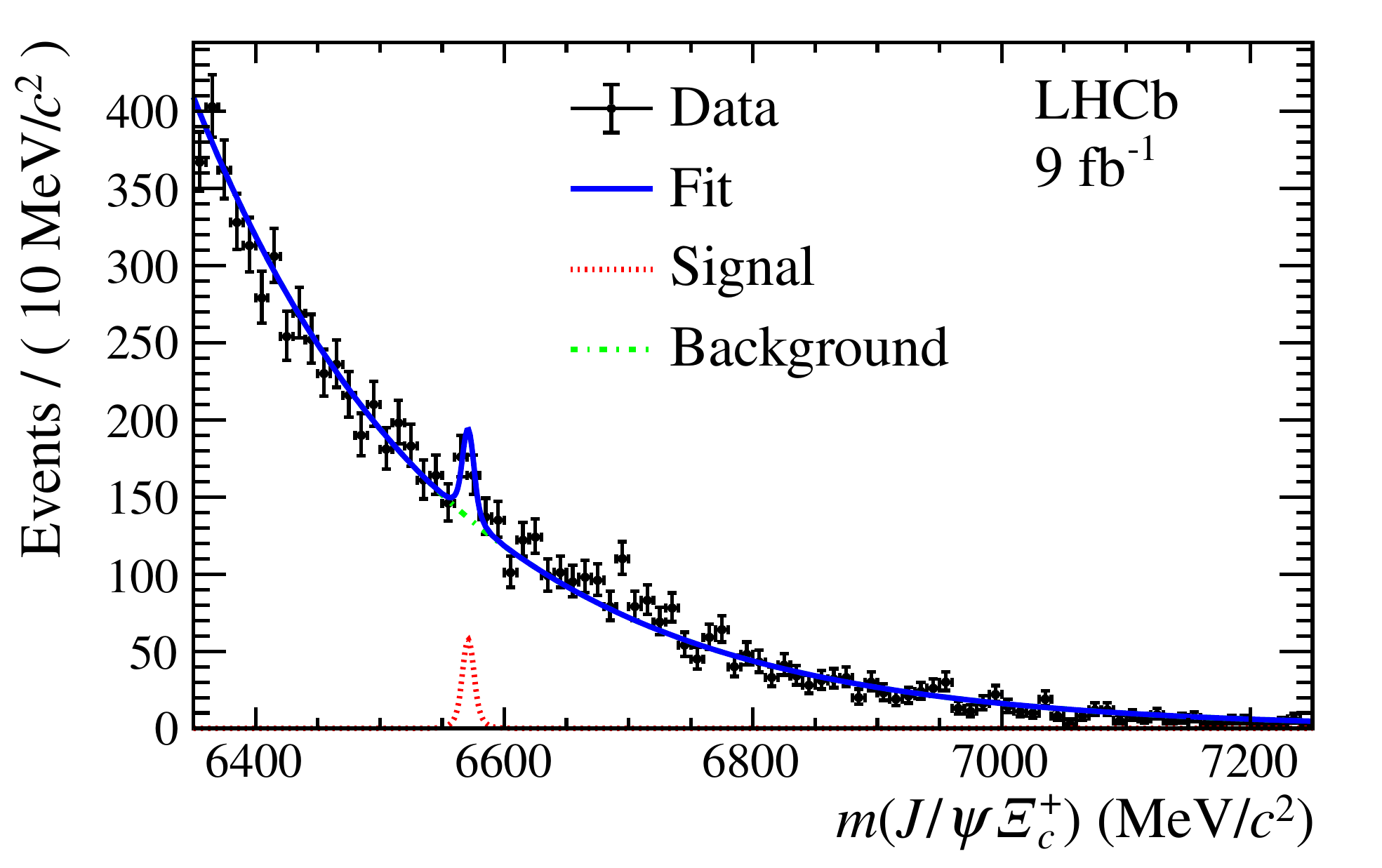}}
   \caption{Mass $m(\jpsi\Xicp)$ distribution of selected $\Xibcp$ candidates for the full data set. The fit (blue solid line) with the largest local significance at the mass of 6571\mevcc is superimposed.}
   \label{fig:xibcmw}
 \end{center} 
\end{figure}

\begin{figure}
  \begin{center}
    \resizebox{0.6\textwidth}{!}{%
      \includegraphics{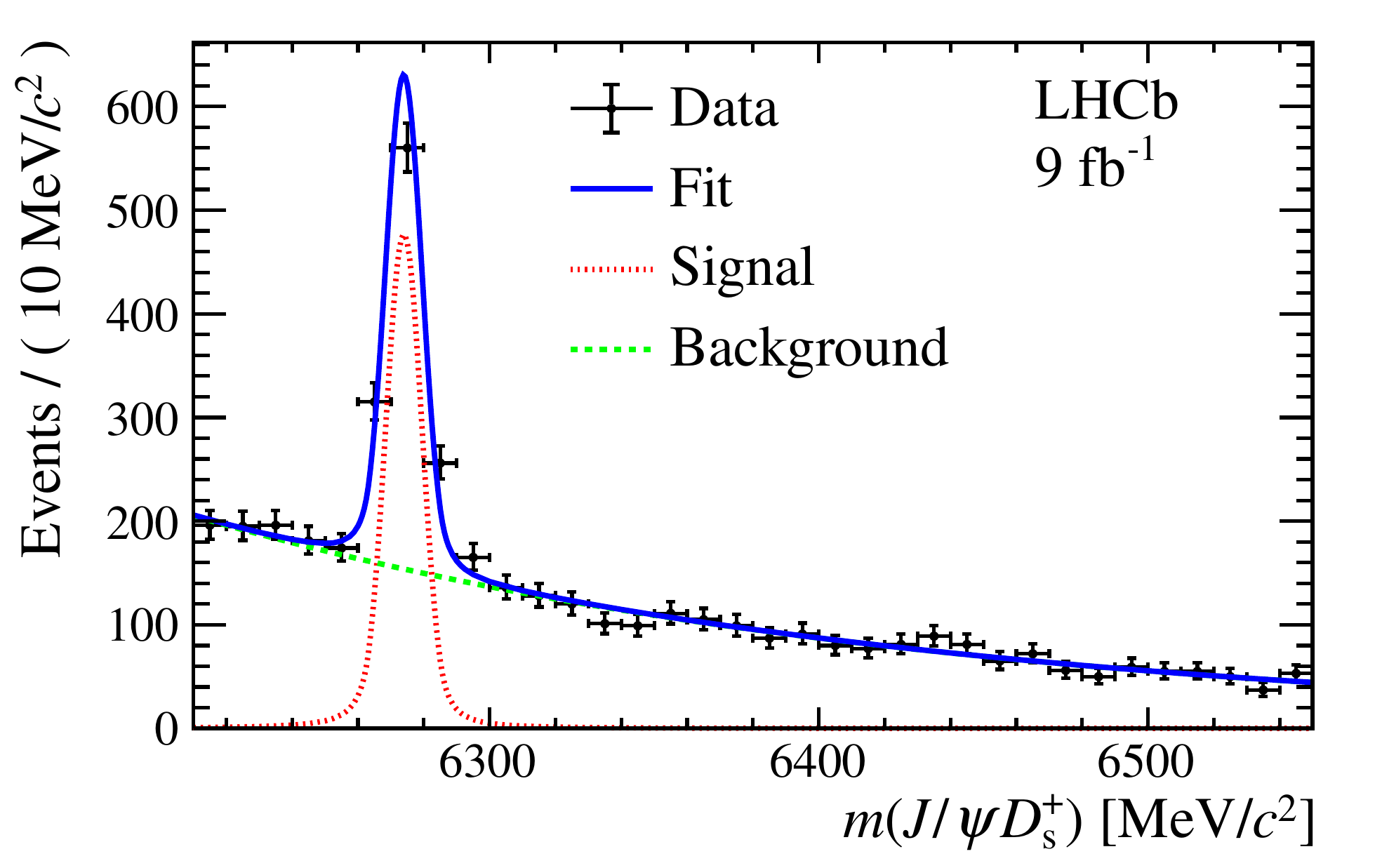}}
   \caption{Mass $m(\jpsi\Dsp)$ distribution of selected $\Bcp$ candidates for the full data set. The fit (blue solid line) is superimposed.}
   \label{fig:Bc_all}
 \end{center} 
\end{figure}

The invariant-mass distributions of selected $\Xibcp$ and $\Bcp$ candidates in the full data sample are shown in Figs.~\ref{fig:xibcmw} and~\ref{fig:Bc_all}, respectively. To improve the mass resolution of the $\Xibcp \;(\Bcp)$ candidates, the $\jpsi\Xicp \;(\jpsi\Dsp)$ invariant mass is calculated by constraining the $\jpsi$ and $\Xicp\; (\Dsp)$ masses to their known values~\cite{PDG2020} and the $\Xibcp\;(\Bcp)$ candidates to originate from their associated PV~\cite{Hulsbergen:2005pu}.

The $\Xibcp$ signal yield is determined from an unbinned maximum-likelihood fit to the $\jpsi\Xicp$ mass distribution. The signal is described by a double-sided Crystal Ball (DSCB) function~\cite{Skwarnicki:1986xj} comprising a Gaussian core with power-law tails on both sides, where the tail parameters depend on the mass resolution, while the combinatorial background is described by an exponential function. 
The dependence of the mass resolution on the \mbox{$\Xibcp$ mass} is determined from simulation.
The mass resolution varies from about 4\mevcc at a $\Xibcp$ mass of 6400\mevcc to 7\mevcc at 7100\mevcc. 
The mass region of interest from 6430 to 7120\mevcc is scanned in 3\mevcc steps, to search for any significant structures.

\begin{figure}
  \begin{center}
    \resizebox{0.6\textwidth}{!}{%
      \includegraphics{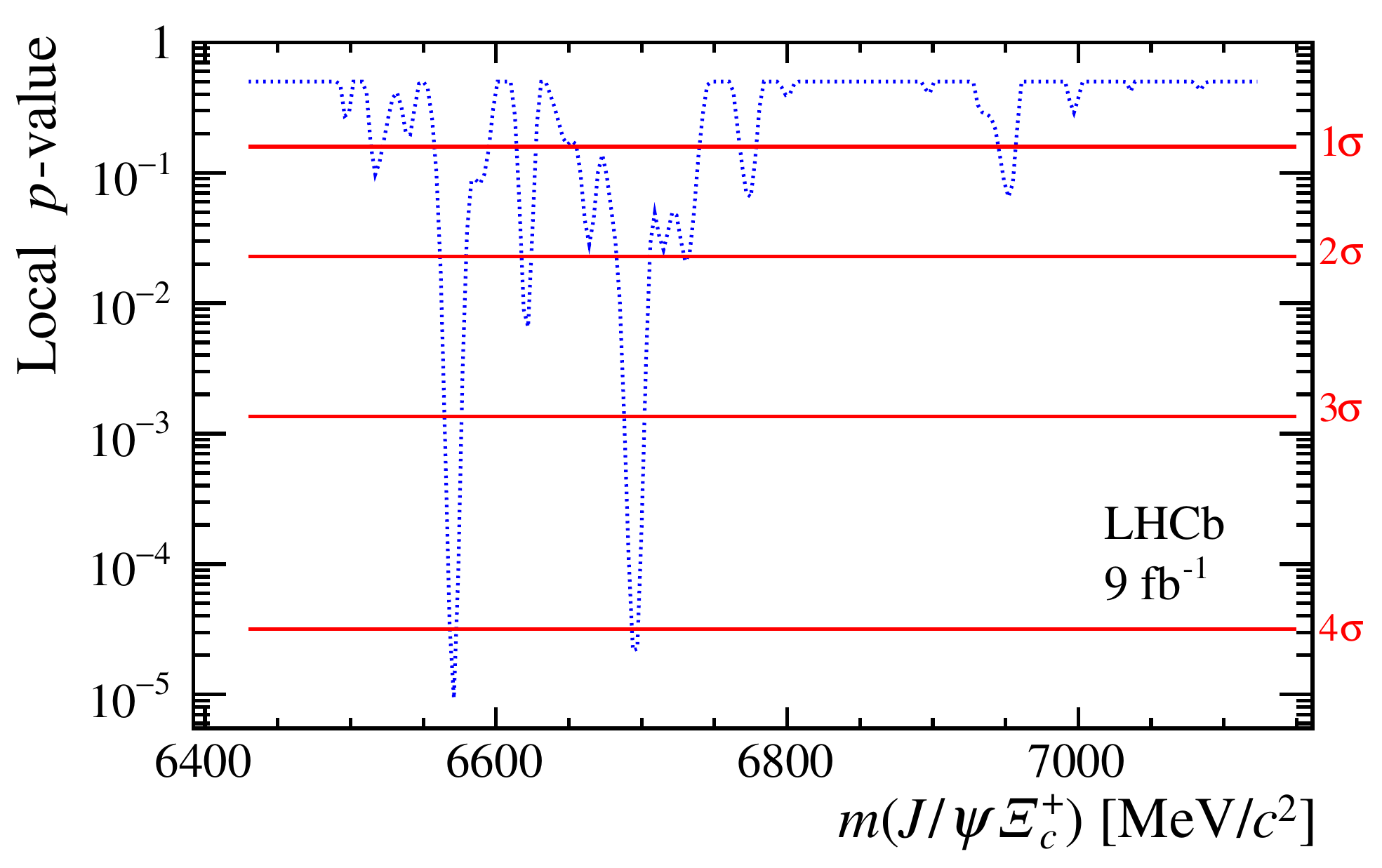}}
   \caption{Local $p$-value in the $m(\jpsi\Xicp)$ range $6430-7120\mevcc$.}
   \label{fig:pvalue}
 \end{center} 
\end{figure}

\begin{figure}
  \begin{center}
    \resizebox{0.6\textwidth}{!}{%
      \includegraphics{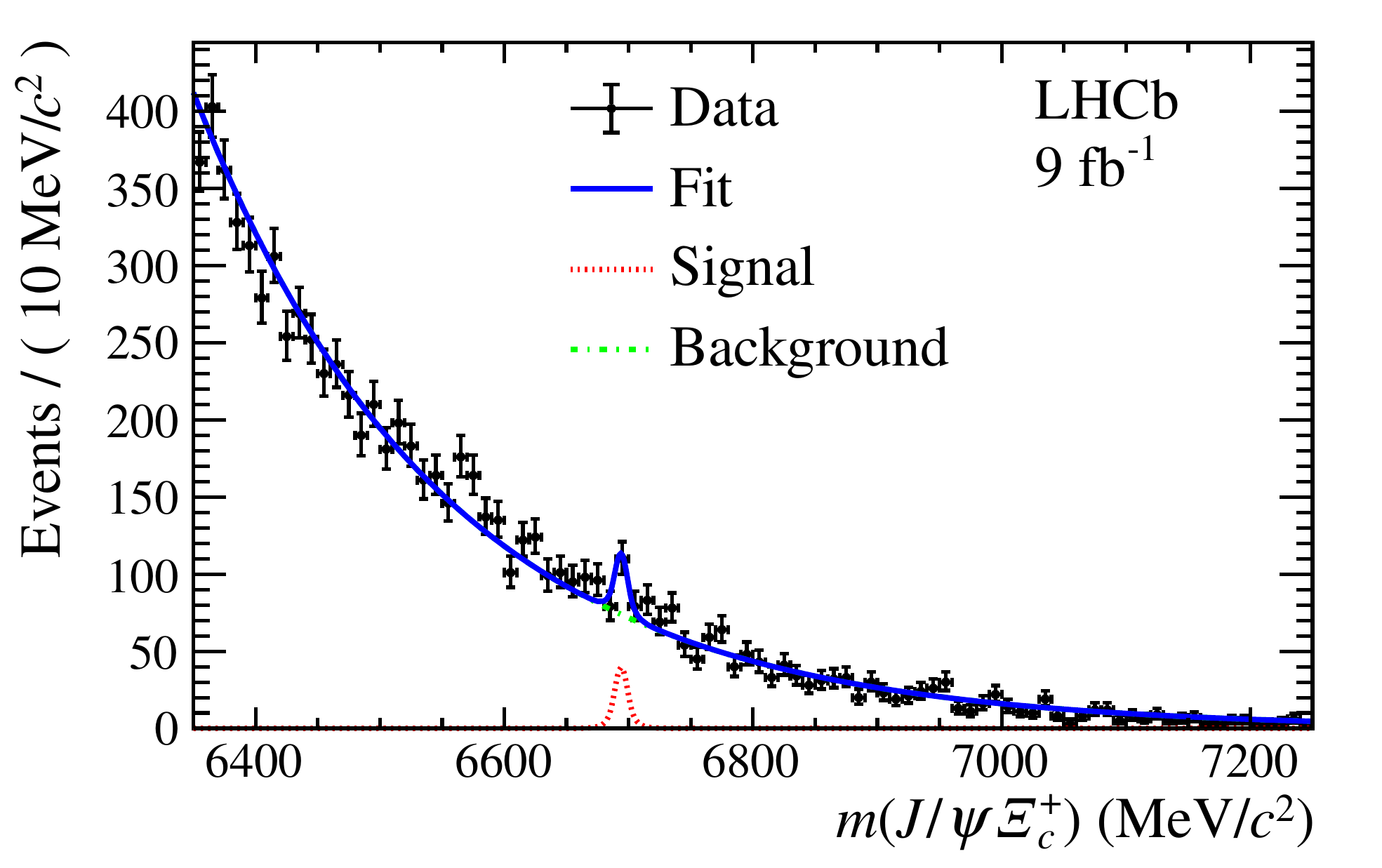}}
   \caption{Mass $m(\jpsi\Xicp)$ distribution of selected $\Xibcp$ candidates for the full data set. The fit (blue solid line) with the second largest local significance at the mass of 6694\mevcc is superimposed.}
   \label{fig:xibc6694}
 \end{center} 
\end{figure}

The local significance of a signal peak is quantified with a $p$-value, which is calculated from the likelihood ratio between the background-plus-signal 
and the background-only hypotheses~\cite{Wilks:1938dza}.
The local $p$-value is plotted in Fig.~\ref{fig:pvalue} as a function of $m(\jpsi\Xicp)$, showing a dip around 6571\mevcc, which has the largest local significance, expressed in number of standard deviations ($\sigma$), corresponding to $4.3\sigma$. Another dip is seen around 6694\mevcc, with a local significance of $4.1\sigma$. 
The fit results for 
the two mass peaks at 6571\mevcc and 6694\mevcc are shown in Figs.~\ref{fig:xibcmw} and~\ref{fig:xibc6694}, and the signal yield is $75 \pm 19$ and $58 \pm 16$, respectively.
The global significance is evaluated using pseudoexperiments, by taking into account the look-elsewhere effect~\cite{Gross:2010qma} in the mass range from 6430\mevcc to 7120\mevcc, and is estimated to be $2.8\sigma$ and $2.4\sigma$ 
for the two mass peaks at 6571\mevcc and 6694\mevcc, respectively. 
As no excess above $3\sigma$ is observed,
upper limits on the production ratios are set for the data samples with centre-of-mass energies of $\sqs=8\tev$ and $13\tev$. 

The $\Bcp$ signal yield is determined from an unbinned maximum-likelihood fit to the $m(\jpsi \Dsp)$ distribution. The $\Bcp$ signal is described by a DSCB function with the tail parameters depending on the mass resolution~\cite{LHCb-PAPER-2020-003},
while the combinatorial background is described by an exponential function. 
The fit to the full data set is shown in Fig.~\ref{fig:Bc_all}. A total of $706\pm38$ $\Bcp\to\jpsi\Ds$ signal decays are selected. The signal yields used in the measurement of ${\mathcal{R}}$ are summarised in Table~\ref{tab:efficiency}. 

\section{Efficiency ratios}
\label{sec:pratio}
The efficiency ratio between the $\Bcp$ and $\Xibcp$ modes, defined as $\eps_{\rm norm}/\eps_{\rm sig}$, is determined from simulation, along with corrections to account for small residual differences between data and simulation. 
The signal efficiency depends upon the assumed mass and lifetime of the $\Xibcp$ baryon. Simulated events are generated with a $\Xibcp$ mass of 6900\mevcc and a lifetime $\tau(\Xibcp) = 400\fs$, labelled here as default. 
The tracking and PID efficiencies for both the signal and normalisation modes are corrected using calibration data samples~\cite{LHCb-DP-2013-002,LHCb-PUB-2016-021,LHCb-DP-2018-001}. The PID efficiency correction is applied by
resampling the distributions of  PID observables   in simulation to match those in data for the variables used in the selection and in the BDT classifier before computing the efficiency.
The efficiency ratio and the single-event sensitivity at the default $\Xibcp$ mass and lifetime are summarised in Table~\ref{tab:efficiency} together with the signal yield of the normalisation mode, used in computing the single-event sensitivity.

\begin{table}
  \caption{Efficiency ratios $\varepsilon_{\rm norm}/\varepsilon_{\rm sig}$ between the normalisation and signal modes, signal yields of the normalisation mode $N_{\rm norm}$, and the single-event sensitivity $\alpha$, for the default mass and lifetime of the $\Xibcp$ baryon, $6900\mevcc$ and $400\fs$, respectively. Uncertainties are statistical only.} 
  \centering
  \begin{tabular}{l c r c}
    \hline
    Data sample & $\varepsilon_{\rm norm}/\varepsilon_{\rm sig}$ & $N_{\rm norm}$& $\alpha$ \\
    \hline
2012 (\sqs=\;\;8\tev) & 1.316 $\pm$ 0.013 & 75 $\pm$ 13 & 0.018 $\pm$ 0.003\\
2016 (\sqs=13\tev) & 1.207 $\pm$ 0.007 & 177 $\pm$ 20& 0.0068 $\pm$ 0.0008\\
2017 (\sqs=13\tev) & 1.202 $\pm$ 0.006 & 193 $\pm$ 20& 0.0062 $\pm$ 0.0006\\
2018 (\sqs=13\tev) & 1.222 $\pm$ 0.006 & 220 $\pm$ 21& 0.0056 $\pm$ 0.0005\\    
\hline
\end{tabular}
\label{tab:efficiency}
\end{table}

The efficiency ratio for other lifetime values are obtained by weighting the simulated events to reproduce lifetime hypotheses from 300 to 500$\fs$ in $50\fs$ steps. An event-by-event weight is calculated as
\begin{equation}
    \label{eq:ltweight}
    w=\frac{(1/\tau)\cdot \exp{(-t/\tau)}}{(1/\tau_0)\cdot \exp{(-t/\tau_0)}},
\end{equation}
where $t$ is the $\Xibcp$ decay time, $\tau$ is the new lifetime and $\tau_0$ is the default lifetime. The total efficiency is found to have a linear dependence on the $\Xibcp$ lifetime. The value and uncertainty in the single-event sensitivity $\alpha$ are provided for each lifetime hypothesis and for each data-taking period (Table~\ref{tab:alpha}). The efficiency could also depend on the $\Xibcp$ baryon mass hypothesis in the simulation, since it affects the kinematic distributions of the decay products. To assess this effect, large samples of simulated events are generated with alternative mass hypotheses in the range 6400--7050$\mevcc$ in $50\mevcc$ steps. 
These samples are used to weight the \pt distributions of the final-state particles in the fully simulated $\Xibcp$ decay to match those of the other mass hypotheses, and the efficiency is then recalculated.
A very small dependence on the $\Xibcp$ mass, a 0.4\% relative variation of the signal efficiency due to this weighting, is observed and considered as a systematic uncertainty.  

\begin{table}
  \caption{Single-event sensitivity $\alpha$ in units of 10$^{-3}$ for different lifetime hypotheses of the $\Xibcp$ baryon for different data taking periods. Uncertainties are due to the limited size of the simulated samples and the statistical uncertainties in the measured $\Bcp$ yields.} 
  \centering
  \begin{tabular}{l c c c c c}
    \hline
    Data sample & 300\fs & 350\fs & 400\fs & 450\fs & 500\fs \\
    \hline
2012 (\sqs=\;\;8\tev) & 22 $\pm$ 4~\, & 20 $\pm$ 3~\, & 18 $\pm$ 3~\, & 16 $\pm$ 2~\, & 15 $\pm$ 2~\,\\
2016 (\sqs=13\tev) & 8.4 $\pm$ 0.9 & 7.5 $\pm$ 0.8 & 6.8 $\pm$ 0.8 & 6.3 $\pm$ 0.7 & 5.9 $\pm$ 0.6\\
2017 (\sqs=13\tev) & 7.7 $\pm$ 0.7 & 6.8 $\pm$ 0.7 & 6.2 $\pm$ 0.6 & 5.7 $\pm$ 0.6 & 5.4 $\pm$ 0.5\\
2018 (\sqs=13\tev) & 6.9 $\pm$ 0.6 & 6.2 $\pm$ 0.6 & 5.6 $\pm$ 0.5 & 5.2 $\pm$ 0.5 & 4.9 $\pm$ 0.4\\    
\hline
\end{tabular}
\label{tab:alpha}
\end{table}

\section{Systematic uncertainties}
\label{sec:sysu}
Systematic uncertainties affecting the measurement of ${\mathcal{R}}$ arise from the PID efficiency corrections, 
the track reconstruction efficiency, 
the difference in the $\Xicp \to p K^{-}\pi^{+}$ Dalitz distribution between data and simulation, 
the variation of the efficiency with respect to the $\Xibcp$ mass, 
the mass resolution used in the fit to the $\Xibcp$ mass spectrum,
and the fit model assumed to evaluate the normalisation yield. 
The total systematic uncertainty is calculated as the quadratic sum of each individual uncertainty presented in Table~\ref{tab:sys3}, assuming no correlation between the contributions. 

The largest systematic uncertainty is due to the PID efficiency correction. There are several sources of systematic uncertainty associated to this correction, mainly due to the limited size of the calibration samples, 
the assumption of no correlations between PID variables of each final state particle,
and limitations in the method used to correct the PID variables. The largest contribution to the PID efficiency correction arises from the comparison between the efficiency obtained with the PID variables resampled assuming no correlations between the PID variables for each final state particle, and an alternative correction method that takes into account such correlations. This alternative method requires corrections in a higher number of dimensions of phase space, and can suffer from statistical fluctuations due to limited size of the calibration samples. This comparison gives a 3.6\% contribution to the PID efficiency correction uncertainty. Summing all the contributions in quadrature, the total systematic uncertainty associated to the PID efficiency correction is 4\%.

Two sources contribute to the systematic uncertainty associated to the tracking efficiency. The first uncertainty is statistical, arising from the limited size of the samples used to derive the efficiency correction. The second is due to hadronic interactions with the detector material~\cite{LHCb-DP-2013-002}.
Considering that two out of the three final-state hadrons in the signal and the normalisation modes are common, the effect cancels out in the ratio except for the proton coming from the $\Xicp$ decay and the positively charged kaon from the $\Dsp$ decay.
These two uncertainties are added in quadrature and the total systematic uncertainty due to the tracking efficiency is 0.8\%.

The uncertainty contribution due to the $\Xicp \to p K^- \pi^+$ decay comes from an imperfect modelling of the Dalitz shape in the simulation. A new signal efficiency is obtained using a weighting technique to match the simulated Dalitz distribution with the one from data, and the resulting difference of 0.5\% is taken as a systematic uncertainty.

As described earlier, the $\Xibcp$ selection efficiency is found to depend on the $\Xibcp$ mass at a level of 0.4\%, which is neglected in the efficiency ratio, and is taken as a systematic uncertainty. 
The uncertainty coming from possible variations of the mass resolution, used in the $m(\jpsi\Xicp)$ fit, are obtained by varying the mass resolution by $\pm$10\%. The largest difference between the local significance from the $p$-value scan obtained with different mass resolutions, 1.5\%, is taken as a systematic uncertainty.
The signal yield of the normalisation mode is affected by the fit model. This is evaluated by considering a sum of two Gaussian functions with the same mean but different resolutions, rather than the default DSCB function. The difference between the two measured yields, 0.2\%, is taken as a systematic uncertainty. The total systematic uncertainty on the measurement of the production ratio $\mathcal{R}$ is 4.4\%.

\begin{table}
  \caption{Systematic uncertainties on the measurement of the production ratio, $\mathcal{R}$.} 
  \centering
  \begin{tabular}{l l}
    \hline
    Source & $\mathcal{R}$ [\%]\\
    \hline
PID                                                   & 4.0\\
Tracking                                              & 0.8\\
$\Xicp \to p K^{-}\pi^{+}$ Dalitz distribution        & 0.5\\
$\Xibcp$ mass                               & 0.4\\
Mass resolution                             
     & 1.5\\
$\Bcp$ signal shape                        & 0.2\\     
\hline
Total systematic uncertainty                                                & 4.4\\
\hline
\end{tabular}
\label{tab:sys3}
\end{table}

\section{Results and summary}
\label{sec:res}
The search for the $\Xibcp$ baryon shows no significant signal, and therefore upper limits at the 95\% credibility level (CL) are set. The upper limits include the systematic uncertainties, obtained by convolving the likelihood profile $\mathcal{L(R)}$ with a Gaussian distribution whose width is given by the quadratic sum of the uncertainty in the single-event sensitivity and the systematic uncertainties. The likelihood profile $\mathcal{L(R)}$ 
is calculated by setting different $\Xibcp$ mass hypotheses in the fit to the $\jpsi\Xicp$ invariant-mass distribution, with a step of 3\mevcc. The upper limit at 95\% CL is defined as the value of $\mathcal{R}$ at which the integral of the likelihood profile equals 95\% of its total area.

\begin{figure}[!t]
  \begin{center}
    \resizebox{0.6\textwidth}{!}{%
      \includegraphics{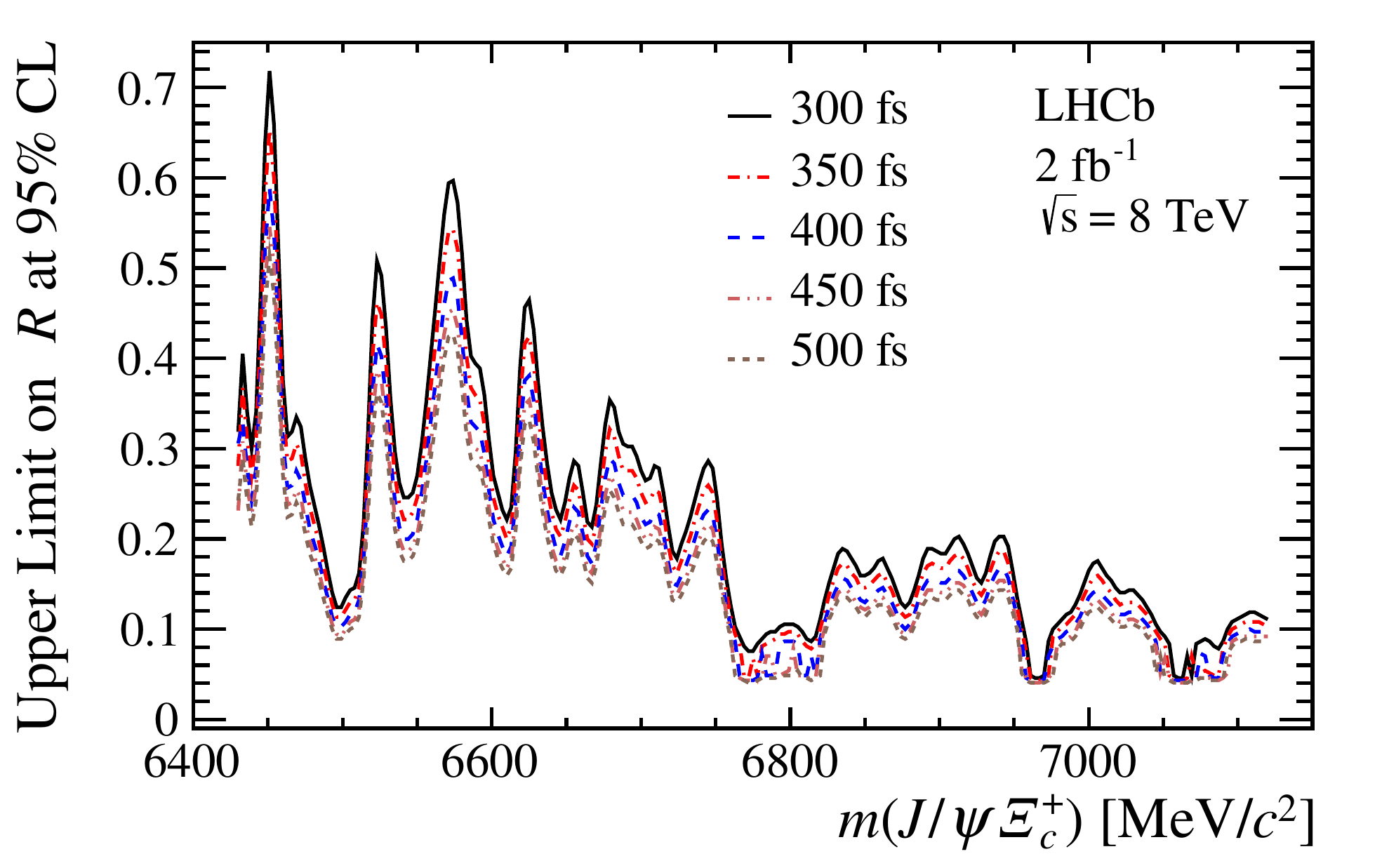}}
   \caption{Upper limits on $\mathcal{R}$ at 95\% CL as a function of $m(\jpsi\Xicp)$ for five $\Xibcp$ lifetime hypotheses (300, 350, 400, 450 and $500\fs$) at a centre-of-mass energy of $\sqs=8\tev$. The curves from top to bottom correspond to lifetime hypotheses from $300\fs$ to $500\fs$, respectively.}
   \label{fig:ul2012}
 \end{center} 
  \begin{center}
    \resizebox{0.6\textwidth}{!}{%
      \includegraphics{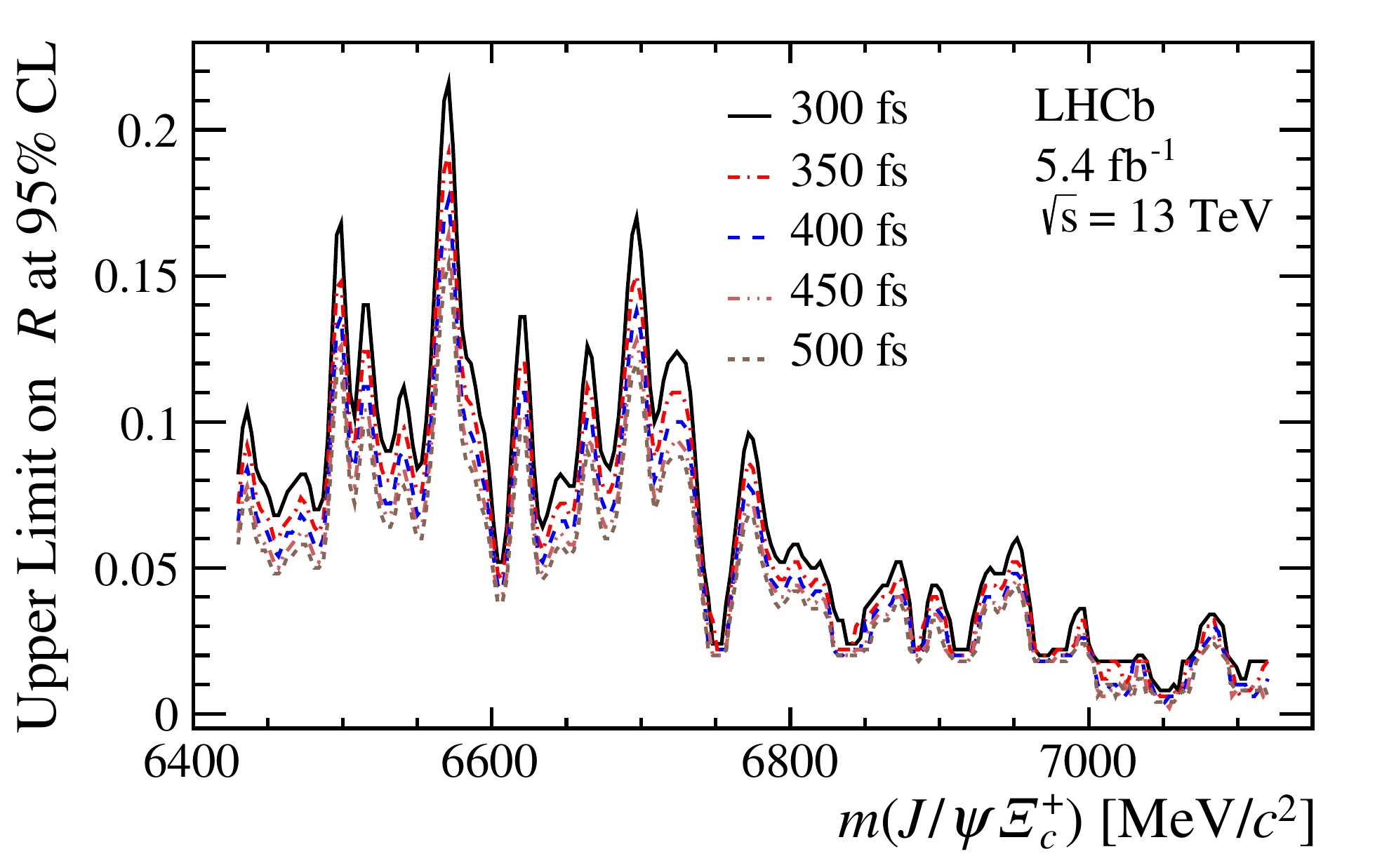}}
   \caption{Upper limits on $\mathcal{R}$ at 95\% CL as a function of $m(\jpsi\Xicp)$ for five $\Xibcp$ lifetime hypotheses (300, 350, 400, 450 and $500\fs$) at a centre-of-mass energy of $\sqs=13\tev$. The curves from top to bottom correspond to lifetime hypotheses from $300\fs$ to $500\fs$, respectively.}
   \label{fig:ul201618}
 \end{center} 
\end{figure}

The resulting 95\% CL upper limits on ${\mathcal{R}}$ as a function of assumed $\Xibcp$ mass are shown in Figs.~\ref{fig:ul2012} and~\ref{fig:ul201618} for the $\sqs=8$\tev and 13\tev data samples, respectively.
The results are restricted to the kinematic region $2.0 < y < 4.5$ and $0 < \pt < 20\gevc$. Upper limits at several different lifetimes of $\Xibcp$ (300, 350, 400, 450 and 500\fs) are also shown at each centre-of-mass energy. The upper limits are set assuming that the kinematic distributions of the $\Xibcp$ baryon follow those of the $\textsc{GenXicc}$ 2.0 model~\cite{Chang:2009va} and that the $\Xibcp$ baryon decays uniformly within the available phase space.

In summary, a first search for the $\Xibcp$ baryon using $\Xibcp\to\jpsi\Xicp$ decays is reported by the LHCb experiment using a $pp$ collision data sample corresponding to an integrated luminosity of 9\invfb, recorded at centre-of-mass energies of 7, 8, and 13\tev.  The most significant peaks in the mass region considered correspond to local (global) significance of $4.3\sigma$ ($2.8\sigma)$ and $4.1\sigma$ ($2.4\sigma$) at $6571\mevcc$ and $6694\mevcc$. Thus, there is no evidence for the $\Xibcp$ baryon with the current data sample.
The 95\% CL upper limits on the relative production rate ${\mathcal{R}}$ of $\Xibcp$ baryons relative to the $\Bcp$ meson, are reported at $\sqs=8$ and 13\tev in the kinematic region $2.0 < y < 4.5$ and transverse momentum $0 < \pt < 20 \gevc$ for a range of possible $\Xibcp$ lifetimes.

In the region favoured by most theoretical models (6700--7029\mevcc), the limits on ${\mathcal{R}}$ for the 13\tev data sample are close to
the rough estimate of 0.015 discussed previously in Sec.~\ref{sec:intro}.
With the larger data samples anticipated in the future running of the LHCb experiment and the inclusion of additional decay modes there is potential for evidence or discovery of the $\Xibcp$ baryon in the future. 

\section*{Acknowledgements}
%
%
\noindent We express our gratitude to our colleagues in the CERN
accelerator departments for the excellent performance of the LHC. We
thank the technical and administrative staff at the LHCb
institutes.
We acknowledge support from CERN and from the national agencies:
CAPES, CNPq, FAPERJ and FINEP (Brazil); 
MOST and NSFC (China); 
CNRS/IN2P3 (France); 
BMBF, DFG and MPG (Germany); 
INFN (Italy); 
NWO (Netherlands); 
MNiSW and NCN (Poland); 
MEN/IFA (Romania); 
MICINN (Spain); 
SNSF and SER (Switzerland); 
NASU (Ukraine); 
STFC (United Kingdom); 
DOE NP and NSF (USA).
We acknowledge the computing resources that are provided by CERN, IN2P3
(France), KIT and DESY (Germany), INFN (Italy), SURF (Netherlands),
PIC (Spain), GridPP (United Kingdom), 
CSCS (Switzerland), IFIN-HH (Romania), CBPF (Brazil),
Polish WLCG  (Poland) and NERSC (USA).
We are indebted to the communities behind the multiple open-source
software packages on which we depend.
Individual groups or members have received support from
ARC and ARDC (Australia);
Minciencias (Colombia);
AvH Foundation (Germany);
EPLANET, Marie Sk\l{}odowska-Curie Actions and ERC (European Union);
A*MIDEX, ANR, IPhU and Labex P2IO, and R\'{e}gion Auvergne-Rh\^{o}ne-Alpes (France);
Key Research Program of Frontier Sciences of CAS, CAS PIFI, CAS CCEPP, 
Fundamental Research Funds for the Central Universities, 
and Sci. \& Tech. Program of Guangzhou (China);
GVA, XuntaGal, GENCAT and Prog.~Atracci\'on Talento, CM (Spain);
SRC (Sweden);
the Leverhulme Trust, the Royal Society
 and UKRI (United Kingdom).



\addcontentsline{toc}{section}{References}
\bibliographystyle{LHCb}
\bibliography{main,standard,LHCb-PAPER,LHCb-CONF,LHCb-DP,LHCb-TDR,Xibc}

\newpage
\centerline
{\large\bf LHCb collaboration}
\begin
{flushleft}
\small
R.~Aaij$^{32}$\lhcborcid{0000-0003-0533-1952},
A.S.W.~Abdelmotteleb$^{50}$\lhcborcid{0000-0001-7905-0542},
C.~Abellan~Beteta$^{44}$,
F.~Abudin{\'e}n$^{50}$\lhcborcid{0000-0002-6737-3528},
T.~Ackernley$^{54}$\lhcborcid{0000-0002-5951-3498},
B.~Adeva$^{40}$\lhcborcid{0000-0001-9756-3712},
M.~Adinolfi$^{48}$\lhcborcid{0000-0002-1326-1264},
H.~Afsharnia$^{9}$,
C.~Agapopoulou$^{13}$\lhcborcid{0000-0002-2368-0147},
C.A.~Aidala$^{76}$\lhcborcid{0000-0001-9540-4988}
% ...

{\footnotesize \it

% ...
$^{9}$Universit{\'e} Clermont Auvergne, CNRS/IN2P3, LPC, Clermont-Ferrand, France\\
% ...
$^{13}$LPNHE, Sorbonne Universit{\'e}, Paris Diderot Sorbonne Paris Cit{\'e}, CNRS/IN2P3, Paris, France\\
% ...
$^{32}$Nikhef National Institute for Subatomic Physics, Amsterdam, Netherlands\\
% ...
$^{40}$Instituto Galego de F{\'\i}sica de Altas Enerx{\'\i}as (IGFAE), Universidade de Santiago de Compostela, Santiago de Compostela, Spain\\
% ...
$^{44}$Physik-Institut, Universit{\"a}t Z{\"u}rich, Z{\"u}rich, Switzerland\\
% ...
$^{48}$H.H. Wills Physics Laboratory, University of Bristol, Bristol, United Kingdom\\
% ...
$^{50}$Department of Physics, University of Warwick, Coventry, United Kingdom\\
% ...
$^{54}$Oliver Lodge Laboratory, University of Liverpool, Liverpool, United Kingdom\\
% ...
$^{62}$Syracuse University, Syracuse, NY, United States\\
% ...
$^{76}$University of Michigan, Ann Arbor, MI, United States, associated to $^{62}$\\
\bigskip
% ...
}
\end{flushleft}

\end{document}